\documentclass[prb,twocolumn,aps,longbibliography]{revtex4-2}
\usepackage{graphicx,amsmath,xcolor}
\usepackage{hyperref}
\usepackage{subcaption}
\usepackage{float}
\usepackage{svg}
\usepackage{enumerate}
\usepackage{silence}
\usepackage{hyperref}
\usepackage{mathtools}
\usepackage[T1]{fontenc}
\DeclarePairedDelimiter{\bra}{\langle}{\rvert}%
\DeclarePairedDelimiter{\ket}{\lvert}{\rangle}%
\DeclarePairedDelimiterX\innerp[2]{\langle}{\rangle}{#1\delimsize\vert\mathopen{}#2}%
\DeclarePairedDelimiterX\braket[2]{\langle}{\rangle}{#1\delimsize\vert\mathopen{}#2}%
\DeclarePairedDelimiterX\braketOP[3]{\langle}{\rangle}{#1\,\delimsize\vert\,\mathopen{}#2\,\delimsize\vert\,\mathopen{}#3}%
\DeclarePairedDelimiterX\ketbra[2]{\lvert}{\rvert}{#1\delimsize\rangle\!\delimsize\langle#2}%
\DeclarePairedDelimiterX\outerp[2]{\lvert}{\rvert}{#1\delimsize\rangle\!\delimsize\langle#2}%
\DeclarePairedDelimiterX\projector[1]{\lvert}{\rvert}{#1\delimsize\rangle\!\delimsize\langle#1}%
\newcommand{\comment}[1]{}
\newcommand{\subsecref}[1]{S-\Roman{section}.\arabic{subsection}}

\begin{document}
    \title{Large Scale Optimization of disordered Hubbard models through Tensor and Neural Networks}
    \author{Jacob R. Taylor}
    \author{Sankar Das Sarma}
    \affiliation{Condensed Matter Theory Center and Joint Quantum Institute, Department of Physics, University of Maryland, College Park, Maryland 20742-4111 USA}
\begin{abstract}
{We theoretically demonstrate a practical method for tuning randomly disordered 2D quantum-dot grids underlying spin qubit platforms using vision-based neural networks trained on tensor-network generated charge-stability data. We show that a simulatable local $3\times 3$ window already contains sufficient information to tune the central dot within a much larger array, thereby validating a sliding-window approach in which one tunes a local region and then translates that window across the lattice to calibrate a larger device. This avoids the computationally intractable necessity for obtaining the ground states for large systems with exponentially large Hilbert space. For the experimentally relevant case where only the on-site disorder is unknown, the neural network predicts the relevant parameters with very high fidelity in the $3\times 3$ setting [$R^2 >0.99$], and after fine tuning on only a small number of larger-device samples, it retains high accuracy for the central dot of a $5\times 5$ plaquette [$R^2\approx 0.98$]. When all the dots parameters are treated as unknown, prediction of the on-site disorder remains robust [$R^2>0.9$ for both $3\times 3$ and $5\times 5$], although the remaining parameters are substantially more difficult to infer from the same charge-stability data. This shows that the most practically important disorder parameter for tuning can still be inferred reliably even in the fully disordered setting for the computationally difficult 5x5 arrays. Our results therefore show that accurate tuning of large 2D quantum-dot devices can be achieved from simulations of only a limited local window, rather than from modeling the full array or a particularly large grid. While modest increases in window size should further improve accuracy by incorporating more of the nearby environment, we show that tensor-network simulations of relatively small windows are already sufficient in practice. Since the quantum spin qubits are modeled by a disordered Hubbard model, our work establishes the possibility of using neural networks for studying the 2D Hubbard model.}
\end{abstract}
\maketitle 
\textit{Introduction.\textemdash} Machine learning has emerged as a promising tool for investigating, automating control, and mitigating disorder in quantum hardwares, particularly in regimes where a sufficiently accurate physical model exists but experimental devices still contain unknown disorder in their parameters.\cite{taylor2025neural,durrer2020automated,moon2020machine,taylor2023machine,taylor2025mitigating,taylor2026predicting,zwolak2024data,rao2025modular,buterakos2026qdflow}. This is especially true in semiconductor quantum-dot spin-qubit platforms, which have emerged as promising qubit candidates  \cite{hanson2007spins,sarma2001spin,van2002electron,sarma2005spin,chatterjee2021semiconductor,burkard2023semiconductor}. In these systems, the disordered extended Hubbard model captures the underlying physics with high accuracy \cite{barthelemy2013quantum,byrnes2008quantum,hensgens2017quantum,stafford1994collective, Kotlyar1998,kotlyar1998correlated,Chou2026}, yet the precise device parameters are generally unknown because of disorder, which is invariably present in the experimental systems. As a result, qubit operation requires continual parameter tuning for disorder/drift and remains a central problem in semiconductor quantum dot platforms \cite{zwolak2023data,zwolak2023colloquium,hensgens2017quantum}. Many approaches to this problem have been considered \cite{durrer2020automated,wang2023automated,schuff2024fully,taylor2025neural}. However, most focus only on tuning a small number of dots at a time.

In our previous work \cite{taylor2025neural}, we showed that neural networks can infer Hubbard parameters in small, though larger than previously studied, quantum-dot systems directly from charge stability diagram data, providing a route toward automated control in experimentally relevant settings since arbitrarily large amount of charge stability data can be easily experimentally obtained by changing the system parameters in order to train the neural network. Here, we consider the substantially more challenging case of larger two-dimensional quantum-dot arrays, where scaling is nontrivial for both computational and physical reasons. The idea is to use simulated training data on realistic quantum dot models to train a neural network which produces arbitrarily high fidelity on the test data so as to establish the efficacy and the reliability of this approach for future use on real experimental data. On the computational side, the Hilbert space grows exponentially with system size, making exact diagonalization (to obtain the training data) impractical beyond relatively small devices. On the physical side, the measured charge-stability response of a given dot depends not only on its own local parameters but also on neighboring dots \cite{wang2011quantum}. This raises a central question: whether experimentally accessible measurements from a restricted local tuning region retain enough information to permit reliable inference of the local microscopic parameters. It is not obvious at all that this is possible even as a matter of principle since the underlying many-body Hamiltonian cannot be solved for an arbitrarily large system. To address this key issue, we move beyond the exact-diagonalization regime and instead use tensor-network methods to simulate the measurement data needed for machine-learning based tuning of 2D Hubbard-model devices. In particular, we use matrix product state (MPS) and density matrix renormalization group (DMRG) \cite{white1992density,schollwock2011density} techniques to study much larger quantum-dot lattices of sizes such as $3\times 3$ and $5\times 5$. Although tensor-network methods do not eliminate the exponential complexity of the full problem, they substantially expand the range of device sizes that can be simulated in practice, especially in the locally focused setting relevant to the present work. Figuring out whether such local measurements contain sufficient information for parameter inference in a realistic quantum dot setting is essential for determining whether automated AI-inspired tuning strategies can remain effective as quantum-dot architectures grow in size.

Our central goal is this, to determine whether local measurements from a finite tuning window contain enough information to infer the Hubbard parameters for the theoretical model underlying the 2D coupled quantum fit system, and thus to tune, a central dot within a larger lattice. To do this, we combine training data generated from the tensor network with a vision-based neural network \cite{krizhevsky2012imagenet,he2016deep,dosovitskiy2020image} that takes input from charge-stability diagrams and predicts the unknown parameters of the underlying Hubbard-model. This provides a practical route for studying larger and more realistic quantum-dot devices within the extended Hubbard model. The local-window approach is motivated by the physical locality of the problem: rather than attempting to tune an entire large array at once, one can tune a central region, then slide the window across the device to gradually calibrate the full lattice. In this sense, the present work is aimed at establishing the feasibility of scalable automated tuning for large quantum-dot arrays.

We show that the proposed local sliding-window strategy remains highly effective beyond the small-system regime. In the case where only the on-site (ie on-dot) energy ($\epsilon_i$) is unknown, predictions are highly accurate in $3\times 3$ windows and remain so in $5\times 5$ devices after fine-tuning on only a small number of larger-device samples. We mention that $\epsilon_i$ is the key parameter in determining the charge stability diagram based on capacitance measurements. When all Hubbard parameters are treated as unknown, $\epsilon_i$ can still be inferred robustly for both sizes, although the remaining parameters are much less accessible from the same charge-stability data. Moreover, the residual difference between the $3\times 3$ and $5\times 5$ predictions appears to be approximately linear, suggesting that transfer to larger systems may require only a modest calibration step which should be even achievable for the experimental data. Overall, this provides strong evidence for the scalability of the sliding-window approach to larger quantum-dot arrays.


\textit{Physical Model.\textemdash}
We consider a 2D grid of quantum dots within semiconductor platforms such as silicon heterostructures, which can be accurately modeled by the generic Hubbard model \cite{barthelemy2013quantum,byrnes2008quantum,hensgens2017quantum}. The generic extended Hubbard model is expressed as follows:
\begin{multline}
H=-\sum_{<i,j>, \sigma}t_{ij} \left(c^\dagger_{i\sigma}c_{j\sigma}+h.c.\right)-\sum_i \epsilon_i n_i\\ +\sum_{<i,j>}V_{ij}n_in_j +\sum_i \frac{U_i}{2} n_i (n_i -1)
\end{multline}

\noindent Here, $n_i=n_{i\uparrow}+n_{i\downarrow}$ is the number operator, and $c_{i\sigma}$ is the fermion annihilation operator, with $i$ enumerating the site locations and $\sigma$ the spin degree of freedom (up/down spins). The parameter $t_{ij}$ represents the hopping amplitude between sites, $\epsilon_i$ is the single-particle energy, $V_{ij}$ denotes the inter-site Coulomb repulsion (where $i,j$ indicate different sites), and $U_i$ stands for the on-site repulsion (essentially the diagonal $V_{ii}$ component of $V_{ij}$). In our problem, the Hubbard model parameters are assumed to be unknown. Only nearest-neighbor interactions on the grid are considered, such that $V_{ij}=t_{ij}=0$ for non-nearest neighbors although generalizations to more complicated models, if necessary, should be straightforward at the cost of having many unknown parameters.

The objective is to determine the Hubbard model parameters of the quantum dot system, namely the set ${t_{ij},V_{ij},U_i,\epsilon_i}$, from experimentally measurable occupations. We treat the system as at zero temperature which even in a large system with small coupling constant, T around $\frac{k_B}{\beta}=0.0026 \langle U\rangle$ (with $\langle U\rangle$ being the average U value over the sites), is experimentally achievable \cite{hensgens2017quantum}, and should be sufficient to maintain an effectively pure state. Again, finite-T generalization is straightforward at the cost of substantial increase in the computational cost.


\textit{Method.\textemdash}
The general method consists of creating a tuning window that can be efficiently simulated, such as \(3\times 3\), and using it to tune the center dot within the window. Information about the center dot is obtained by adjusting the chemical potentials of each neighboring dot and measuring the expectation values of the occupation vector \(\vec{n}(\vec{\mu})=(\langle n_1\rangle,\langle n_2\rangle,\ldots)\) formed from the dots within the window. Each dot \(i\) has chemical potential \(\mu_i=\mu_b+\delta\mu_i\), where \(\mu_b\) is a background chemical potential uniform across all dots and \(\delta\mu_i\) is the dot-specific deviation from that value. The input measurements consist of a series of \(\vec{n}(\vec{\mu})\) values where, iterating over the center dot \(i_c\) and each nearest-neighbor pair \(j_{nn}\), \(\delta\mu_{i_c}=-\delta\mu_{j_{nn}}\) is varied between \([-0.6,0.6]\) with 15 points, and \(\delta\mu_i=0\) for \(i\neq i_c,j_{nn}\). At the same time, \(\mu_b\) is also varied independently for all dots between \([-0.5,2.5]\) with 5 points. For each configuration \(m\) of \(\vec{\delta\mu}^{(m)}\) and \(\mu_b^{(m)}\), or together \(\vec{\mu}^{(m)}\), \(\vec{n}^{(m)}\) is measured, and all such measurements within the window are combined into a single matrix,
\[
X=[\vec{n}^{(m=1)},\vec{n}^{(m=2)}, \ldots].
\]
This matrix \(X\) serves as the input to the neural network both for the theoretical case presented here and for a possible experimental application. It is the charge stability diagram for a particular disorder or device realization. The neural network function \(F_{NN}\) is trained to predict the Hubbard model parameters of all dots within the window,
\[
F_{NN}(X)=[\vec{\epsilon},\vec{U},\vec{V},\vec{t}],
\]
where we generate 6500 different disorder realizations of model parameters sampled randomly, alternating between uniform and Gaussian distributions with equal probability for \(\vec{U}\), \(\vec{V}\), and \(\vec{\epsilon}\), within the ranges \(\epsilon_i \in [-0.5,0.5]\), \(U_i \in [3,5]\), \(V_{ij} \in [0,0.4]\), and \(t_{ij} \in [0.1,0.5]\). We also fix \(t_{12}=0.25\) to set the energy scale.

Although a single network could be trained to predict all Hubbard model parameters, we found that accuracy is higher when a separate network is trained for each parameter, since there is no guarantee that the visual features used to determine each parameter are the same. We also note that, although the network is trained to predict all parameters within the window, this is not the main focus. Rather, the focus is on the center dot, as we intend to slide the window to enable tuning of all dots in a larger grid. 

\textit{Simulations.\textemdash}
The simulations use MPS and DMRG \cite{schollwock2011density} in ITensor \cite{itensor} to model \(3\times 3\) and \(5\times 5\) quantum-dot grids. Since an MPS is intrinsically one-dimensional, the 2D lattice is mapped onto a 1D ordering by snaking through the sites (Fig.~\ref{fig:TensorNetwork}). For each configuration of \(\vec{\mu}^{(m)}\) in a sample device, we perform DMRG to obtain the ground state and then compute the site occupations \(\langle n_i\rangle\), yielding \(\vec{n}(\vec{\mu})\). Due to computational limitations, we use 5 DMRG iterations at the first point of the charge stability diagram and then 5 more at each subsequent point, initialized from the previous state. The bond dimension is capped at 100, which was sufficient here, though likely too small for experimental applications.

We note that our focus is only on local site expectation values, rather than more intricate observables such as correlations, so the calculations should converge rapidly even at relatively low bond dimension. For more computationally demanding applications, such as larger lattices, it may also be advantageous to pre-train using low-bond-dimension data and then fine-tune on a smaller set of high-bond-dimension samples. We further note that increasing the bond dimension generally improves neural network fidelity, since a less converged state effectively introduces an additional source of disorder from the network’s perspective.

Due to computational limitations, we consider only fixed boundary conditions and zero temperature (for the corresponding finite-temperature procedure, see SM-\ref{sec:nonzerotemp}). The simulation procedure consists of sweeping the Hamiltonian parameters and performing DMRG at each point without quantum-number restrictions. Although the Hamiltonian preserves quantum numbers, the ground-state sector changes during the parameter sweep, and these changes provide the information used by the neural network.

\comment{
\section{Training}
-The training process in general makes use of two different processes, early s

-In particular we only generate charge stability diagrams relevant for connections between the center dot, 

-In both cases charge stability diagrams are only generated for the 3x3 window of whatever the full lattice size is.  We focus on tuning the center dot within a larger array with the intention that in practice the window will slide to allow tuning a large grid. 

In general the hybrid process consists of first training the network for the 3x3 lattice data, 

-We make use of a very low point dense charge stability diagram 

-Discuss the hybrid process to allow tuning the multi-scales.
}
\begin{figure}[h]
    \centering
    \begin{subfigure}{0.30\textwidth}
        \centering
        \includegraphics[width=\textwidth]{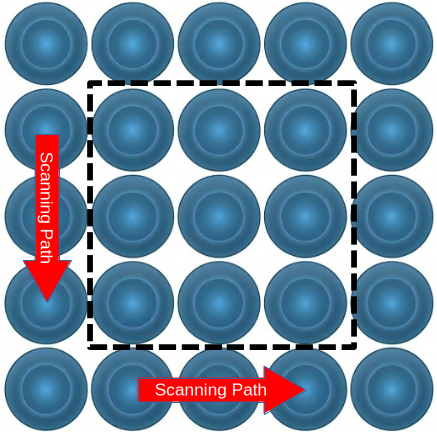}
    \end{subfigure}
    \caption{Sliding-window process diagram. The process for tuning a larger lattice of dots consists of training on an easily simulated window, such as \(3\times 3\), and including a small number of additional \(5\times 5\) samples for fine-tuning on larger arrays of dots. The goal is to tune the center dot of the window and then slide the window across the lattice to tune the entire array, potentially incorporating information from previous tuning operations into future models.}\label{fig:SlidingWindow}
\end{figure}

\textit{Neural Network.\textemdash}
The neural network used in this work is a simple convolutional architecture composed of 3D convolutional layers, 3D max-pooling layers, and squeeze-and-excitation layers \cite{hu2018squeeze}. The network comprises three stages. The first stage is an initial processing block in which the input tensor of size \((9,4,5,15)\) is passed through a 256-filter 3D convolutional layer, followed by a 3D max-pooling layer with kernel size 2 and a second 256-filter 3D convolutional layer. This is followed by two further processing blocks, each consisting of a 3D convolutional layer, a squeeze-and-excitation layer, and a pooling layer, with 256 and 512 filters, respectively. The squeeze-and-excitation layers \cite{hu2018squeeze} introduce global information into the otherwise purely convolutional architecture by reweighting channels according to the input and thereby emphasizing the most relevant features. In the final stage, an adaptive pooling layer contracts the remaining spatial dimensions, after which the representation is processed by a dense layer with 512 nodes and a linear output layer. The output dimensionality depends on the target parameters predicted within the window: all lines on the center dot (4), the onsite values for the center dot and its neighbors (5), or only the onsite value of the center dot (1), although the last option was not used during training.

This architecture was found to outperform the alternatives considered. In particular, a standard vision transformer \cite{dosovitskiy2020image} was initially tested, but transfer learning from the \(3\times 3\) to the \(5\times 5\) case proved substantially less effective, motivating the adoption of a convolutional model. A likely explanation is that the vision transformer’s greater long-range expressivity led to overfitting on the \(3\times 3\) training data. By contrast, the simpler convolutional architecture, with its more limited connectivity, appears better able to capture features that remain consistent across devices of different sizes. Similarly, although residual connections often improve convolutional networks \cite{he2016deep}, they produced at most marginal gains here and were therefore omitted in favor of a simpler model. This may reflect some functional overlap with the squeeze-and-excitation layers.

Training was performed for up to 1000 epochs, with early stopping typically occurring after approximately 200 epochs, once validation performance ceased to improve. The learning rate was gradually reduced from \(1\times 10^{-3}\) to \(5\times 10^{-5}\), and the weight decay from \(10^{-4}\) to \(10^{-5}\), with both schedules determined by the onset of performance plateaus on the validation set.

\begin{figure}[h]
    \centering
    \begin{subfigure}{0.49\textwidth}
        \centering
        \includegraphics[width=\textwidth]{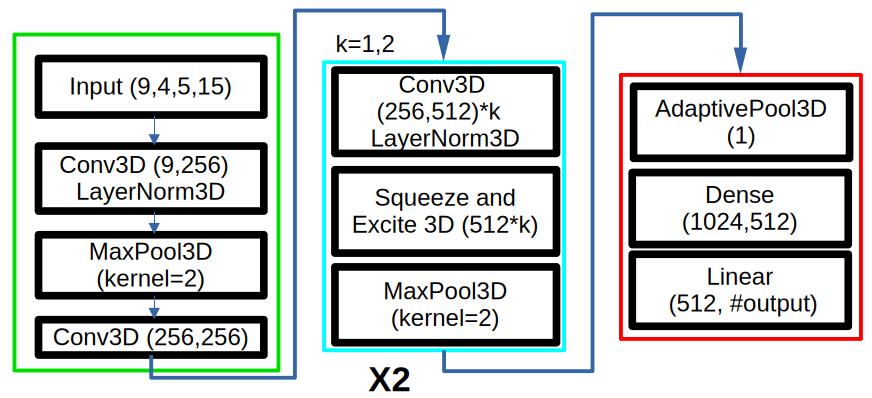}
    \end{subfigure}
    \caption{Neural network diagram. The network is divided into three stages: an input stage (green), two processing blocks (blue), and an output stage (orange). The input stage applies two 3D convolutional layers with layer normalization, separated by 3D max pooling. Each processing block (labeled k=1,2) contains a 3D convolutional layer with layer normalization, a squeeze-and-excitation layer, and 3D max pooling with kernel size 2. The output stage uses adaptive 3D pooling to reduce the remaining spatial dimensions to one, followed by a dense layer and a linear output layer.}\label{fig:NeuralNetwork}
\end{figure}

\textit{Results.\textemdash}
We begin by considering the problem of predicting parameters within the \(3\times 3\) grid using output from tensor-network simulations. We first consider the case in which only \(\epsilon_i\) disorder is present, meaning that all other parameters are assumed known and the task is to predict \(\epsilon_i\). As expected from our previous work \cite{taylor2023machine}, the neural network is able to successfully predict the \(\epsilon_i\) parameters for the nearest neighbors of the center dot within the \(3\times 3\) window, achieving a high fidelity of \(R^2=0.9916\) and a standard-deviation error (root of mean squared error) of \(\Delta\epsilon_i=0.037\). We note that this error corresponds to predicting the disorder for all neighbors of the center dot simultaneously, despite using only the \(\delta\mu_i\) deviations for center-dot nearest-neighbor pairs. \comment{For the center dot in particular, we had \ldots\ .} While further improvement can be obtained by feeding all nearest-neighbor pairs into the neural network, we found this computationally too expensive for generating our training dataset. Those with greater computational resources would likely benefit from doing so.

\begin{figure}[h]
    \centering
    \begin{subfigure}[t]{0.49\linewidth}
    \centering
    \captionsetup{justification=raggedright,singlelinecheck=off,skip=-4cm,margin={-0.05cm,0cm}}
    \centering\includegraphics[width=\textwidth]{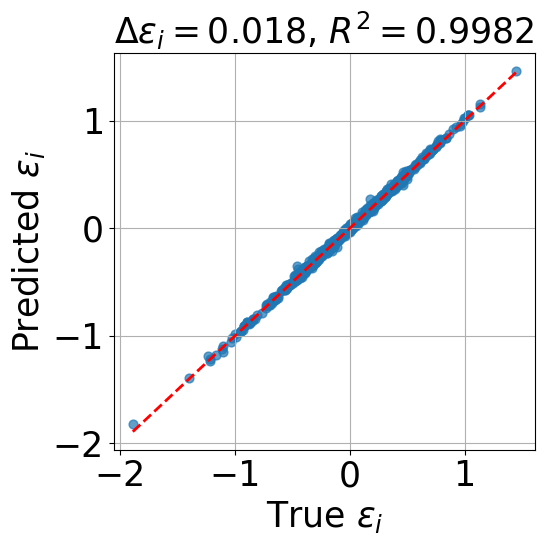}
        \caption{}
    \end{subfigure}
    \hfill
    \begin{subfigure}[t]{0.49\linewidth}
        \centering
        \captionsetup{justification=raggedright,singlelinecheck=off,skip=-4cm,margin={-0.05cm,0cm}}
        \includegraphics[width=\textwidth]{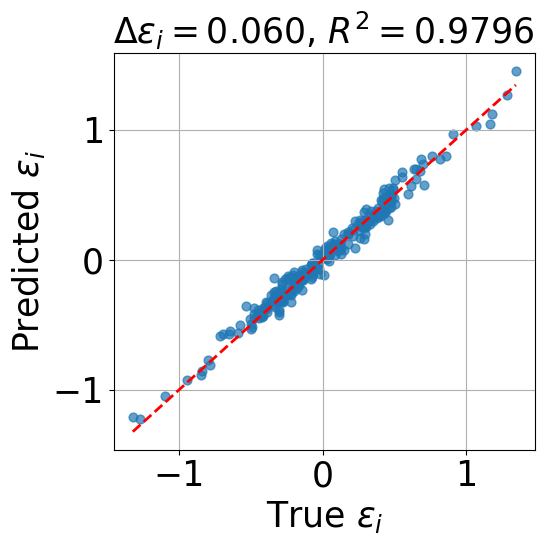}
        
        \caption{}
    \end{subfigure}
    \caption{Predicted vs. expected $\epsilon_i$ for lattice grids of size (a) $3\times 3$ and (b) $5\times 5$. These results correspond to the case in which only the on-site potential at all sites is treated as unknown.}\label{fig:Epsilon}
\end{figure}

With this premise in mind, we seek to tune the center dot both for a \(3\times 3\) grid of dots and for larger grids such as \(5\times 5\). The training process consists of two parts: first, training a network for prediction on the \(3\times 3\) grid using 6500 samples (10\% withheld for testing), and second, by fine-tuning a hybrid network that additionally includes a small number of \(5\times 5\) samples, namely 200 (additional 200 samples used for testing). For the \(5\times 5\) setup, the neural network is given only data from the central \(3\times 3\) window, with the goal of tuning only the center dot. During training, we test the network on both system sizes simultaneously in order to determine the optimal point for early stopping, since otherwise the small number of \(5\times 5\) samples would lead to overfitting. For the \(5\times 5\) test data, we achieve high prediction fidelity for the center dot, with errors of \(\Delta\epsilon_i=0.060\) and \(R^2(\epsilon_i)=0.9796\). Results for both the \(3\times 3\) and \(5\times 5\) cases are shown in Fig.~\ref{fig:Epsilon}.

We found it necessary to include a small number of \(5\times 5\) samples because, without them, the network still appears to predict the relative positions of \(\epsilon_i\), but with a residual discrepancy that may be approximately linear. (See Fig. \ref{fig:NoFineTuning}) In principle, this discrepancy could be calibrated experimentally even for very large systems. Similarly, we expect the effect of increasing system size on the charge stability diagram to diminish for larger and larger systems, given that only a small number of \(5\times 5\) samples is required to learn to achieve high fidelity on the larger system.

This provides strong evidence for the feasibility of our proposed scheme, namely using a sliding window to tune the center dot one site at a time. Using tensor networks, we show that the window size can be extended to \(5\times 5\), and that those with greater computational resources could instead train initially on \(5\times 5\). More generally, we expect that the window need only be large enough to capture the dominant local physics, with larger system sizes entering as corrections that require progressively fewer samples. Beyond some size, we expect these corrections to become effectively noise from the perspective of the neural network and therefore negligible.

While \(\epsilon_i\) is certainly the most important source of disorder, it is also necessary to consider the case in which all model parameters are unknown in order to address two questions: can \(\epsilon_i\) still be predicted for both system sizes when the remaining parameters are also unknown, and if so, what can be said about the other Hubbard parameters? While many tuning tasks may be accomplished with only the first, full tuning requires both.

When we consider the \(3\times 3\) case with all Hubbard parameters unknown, we find that \(\epsilon_i\) can still be predicted with \(\Delta\epsilon_i=0.106\) and \(R^2=0.9284\), using 6500 samples. The main limitation on fidelity is likely the relatively low point density of the charge stability diagrams, together with the relatively small number of training samples. Because we sample on a grid of chemical potentials, the network has difficulty resolving the precise locations of the phase boundaries that it primarily uses for prediction. Nevertheless, the fidelity remains high, and the network appears robust in predicting \(\epsilon_i\) even when all parameters are unknown. When we include the additional 200 samples for the \(5\times 5\) case, we obtain a final fidelity of \(\Delta\epsilon_i=0.121\) and \(R^2=0.9108\). This indicates that having the other parameters unknown does not prevent accurate prediction of \(\epsilon_i\). 
We similarly tested the network on the other Hubbard model parameters, training separate models for each parameter on the \(3\times 3\) and fine-tuned \(5\times 5\) cases using the same dataset for \(\epsilon_i\). We find that \(\epsilon_i\) is the easiest parameter to predict, which is encouraging since it is also the most important for tuning. For the other parameters, however, there is a significant decrease in fidelity compared with our prior work. We expect that these results (including for $\epsilon_i$) could be further improved with greater computational resources, for example by using higher-resolution charge stability diagrams.  
We present a full table of results Tab. \ref{tab:hubbard_table}: 
\begin{table}[h]
\centering

\begin{tabular}{c| c |c |c|c|c}
Parameter & Range & $\sqrt{MSE}_{5\times5}$ &$\sqrt{MSE}_{3\times3}$& $R^2_{5\times5}$ &  $R^2_{3\times 3}$\\
\hline
$\epsilon_i$ &[-0.5,0.5]  & 0.121 &0.106& 0.9108 &0.9284\\
$U_i$   & [3,5] & 0.434 &0.438& 0.6873& 0.6953\\
$V_{ij}$ & [0,0.4] &  0.083&0.081& 0.6731 &0.6823\\
$t_{ij}$   &[0.1,0.5]  & 0.098  &0.075& 0.5146 &0.7178\\
\end{tabular}
\caption{Parameter ranges and fidelities, measured by root mean squared error ($\sqrt{MSE}$) and $R^2$, for the center dot in a $5\times 5$ and $3\times 3$ grid tuned using a $3\times 3$ charge stability diagram window. For $\epsilon_i$ and $U_i$, fidelities are reported for the center dot itself. For $V_{ij}$ and $t_{ij}$, fidelities are averaged over the four interdot connections to the center dot.}
\label{tab:hubbard_table}
\end{table}

\begin{figure}[h]
    \centering
    \begin{subfigure}[t]{0.49\linewidth}
    \centering
    \captionsetup{justification=raggedright,singlelinecheck=off,skip=-4cm,margin={-0.05cm,0cm}}
    \centering\includegraphics[width=\textwidth]{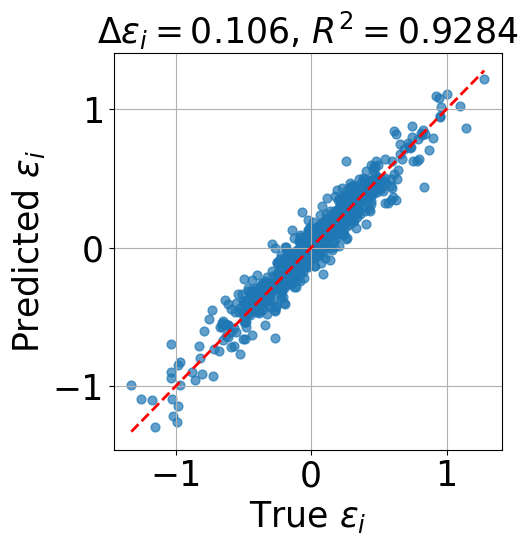}
        \caption{}
    \end{subfigure}
    \hfill
    \begin{subfigure}[t]{0.49\linewidth}
    \captionsetup{justification=raggedright,singlelinecheck=off,skip=-4cm,margin={-0.05cm,0cm}}
        \centering
        \includegraphics[width=\textwidth]{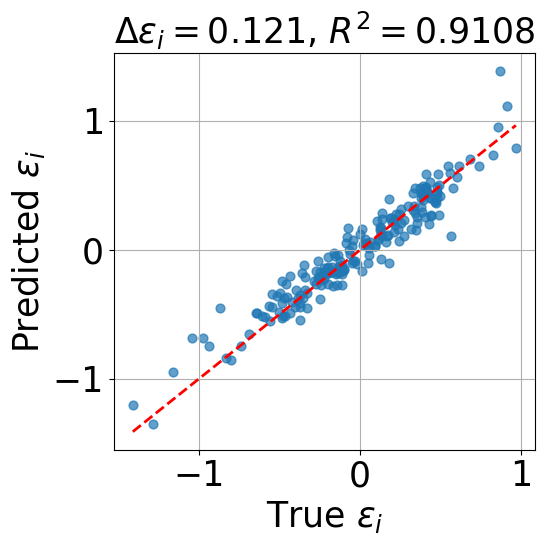}
        \caption{}
    \end{subfigure}
    \caption{Predicted vs. expected $\epsilon_i$ for lattice grids of size (a) $3\times 3$ and (b) $5\times 5$. These results correspond to the case in which all parameters of the Hubbard model are unknown.}\label{fig:AllEpsilon}
\end{figure}

\textit{Conclusion.\textemdash} We have shown that tensor-network simulations combined with machine learning provide a practical route toward scalable automated tuning of large quantum-dot arrays. By using DMRG and MPS methods to simulate local measurement windows, we demonstrate that charge-stability diagram data from a small tuning region contains sufficient information to infer the central dot's model parameters with high fidelity, even in larger lattices. In particular, we find that a network trained primarily on \(3\times 3\) systems and fine-tuned with only a small number of \(5\times 5\) samples can successfully predict the central on-site potential in \(5\times 5\) devices, supporting the viability of a sliding-window tuning strategy for extended arrays. We further find that \(\epsilon_i\), the most important disorder parameter for practical tuning, is also the one most readily predicted by the neural network (and can be done so with high fidelity $R^2>0.9$), while the remaining Hubbard parameters appear substantially more difficult to infer from the same charge-stability data. This limitation is likely due to our computational resources constraining us to use only very low-resolution charge-stability data where transition points are difficult to extract. Taken together, these results provide strong evidence that local-window tuning, supported by tensor-network generated training data, is a promising framework for the calibration and control of large semiconductor quantum-dot devices, and that with improved measurement resolution and additional computational resources, its performance can be extended further. Some details and additional results are provided in the Supplementary Materials.  The numerical techniques are detailed in the references we cite and will be provided on request.

\textit{Acknowledgment.\textemdash } This work is supported by the Laboratory for Physical Sciences. We also thank UMD HPC Zaratan for computational resources provided. 
\bibliography{mainbib}

@article{wang2011quantum,
  title={Quantum theory of the charge-stability diagram of semiconductor double-quantum-dot systems},
  author={Wang, Xin and Yang, Shuo and Sarma, S Das},
  journal={Physical Review B},
  volume={84},
  number={11},
  pages={115301},
  year={2011},
  publisher={APS}
}

@article{hensgens2017quantum,
  title={Quantum simulation of a Fermi--Hubbard model using a semiconductor quantum dot array},
  author={Hensgens, Toivo and Fujita, Takafumi and Janssen, Laurens and Li, Xiao and Van Diepen, CJ and Reichl, Christian and Wegscheider, Werner and Das Sarma, Sankar and Vandersypen, Lieven MK},
  journal={Nature},
  volume={548},
  number={7665},
  pages={70--73},
  year={2017},
  publisher={Nature Publishing Group UK London}
}

@article{chatterjee2021semiconductor,
  title={Semiconductor qubits in practice},
  author={Chatterjee, Anasua and Stevenson, Paul and De Franceschi, Silvano and Morello, Andrea and de Leon, Nathalie P and Kuemmeth, Ferdinand},
  journal={Nature Reviews Physics},
  volume={3},
  number={3},
  pages={157--177},
  year={2021},
  publisher={Nature Publishing Group UK London}
}

@article{taylor2023machine,
  title={Machine learning Majorana nanowire disorder landscape},
  author={Taylor, Jacob R and Sau, Jay D and Sarma, Sankar Das},
  journal={arXiv preprint arXiv:2307.11068},
  year={2023}
}

@article{krizhevsky2012imagenet,
  title={Imagenet classification with deep convolutional neural networks},
  author={Krizhevsky, Alex and Sutskever, Ilya and Hinton, Geoffrey E},
  journal={Advances in neural information processing systems},
  volume={25},
  year={2012}
}

@article{byrnes2008quantum,
  title={Quantum simulation of Fermi-Hubbard models in semiconductor quantum-dot arrays},
  author={Byrnes, Tim and Kim, Na Young and Kusudo, Kenichiro and Yamamoto, Yoshihisa},
  journal={Physical Review B},
  volume={78},
  number={7},
  pages={075320},
  year={2008},
  publisher={APS}
}

@article{barthelemy2013quantum,
  title={Quantum dot systems: a versatile platform for quantum simulations},
  author={Barthelemy, Pierre and Vandersypen, Lieven MK},
  journal={Annalen der Physik},
  volume={525},
  number={10-11},
  pages={808--826},
  year={2013},
  publisher={Wiley Online Library}
}

@article{zwolak2023data,
  title={Data Needs and Challenges of Quantum Dot Devices Automation: Workshop Report},
  author={Zwolak, Justyna P and Taylor, Jacob M and Andrews, Reed and Benson, Jared and Bryant, Garnett and Buterakos, Donovan and Chatterjee, Anasua and Sarma, Sankar Das and Eriksson, Mark A and Greplov{\'a}, Eli{\v{s}}ka and others},
  journal={arXiv preprint arXiv:2312.14322},
  year={2023}
}

@article{burkard2023semiconductor,
  title={Semiconductor spin qubits},
  author={Burkard, Guido and Ladd, Thaddeus D and Pan, Andrew and Nichol, John M and Petta, Jason R},
  journal={Reviews of Modern Physics},
  volume={95},
  number={2},
  pages={025003},
  year={2023},
  publisher={APS}
}

@article{zwolak2023colloquium,
  title={Colloquium: Advances in automation of quantum dot devices control},
  author={Zwolak, Justyna P and Taylor, Jacob M},
  journal={Reviews of modern physics},
  volume={95},
  number={1},
  pages={011006},
  year={2023},
  publisher={APS}
}

@article{hanson2007spins,
  title={Spins in few-electron quantum dots},
  author={Hanson, Ronald and Kouwenhoven, Leo P and Petta, Jason R and Tarucha, Seigo and Vandersypen, Lieven MK},
  journal={Reviews of modern physics},
  volume={79},
  number={4},
  pages={1217},
  year={2007},
  publisher={APS}
}

@article{van2002electron,
  title={Electron transport through double quantum dots},
  author={Van der Wiel, Wilfred G and De Franceschi, Silvano and Elzerman, Jeroen M and Fujisawa, Toshimasa and Tarucha, Seigo and Kouwenhoven, Leo P},
  journal={Reviews of modern physics},
  volume={75},
  number={1},
  pages={1},
  year={2002},
  publisher={APS}
}

@article{kotlyar1998correlated,
  title={Correlated charge polarization in a chain of coupled quantum dots},
  author={Kotlyar, R and Stafford, CA and Sarma, S Das},
  journal={Physical Review B},
  volume={58},
  number={4},
  pages={R1746},
  year={1998},
  publisher={APS}
}

@article{Chou2026,
	title={Spin ladder quantum simulators from spin-orbit coupled quantum dot spin qubits},
	volume={113},
	ISSN={2469-9969},
	url={http://dx.doi.org/10.1103/q9g5-459v},
	number={3},
	journal={Phys. Rev. B},
	publisher={American Physical Society (APS)},
	author={Chou, Yang-Zhi and Das Sarma, Sankar},
	year={2026},
    pages={035124},
	month=jan }

@article{Kotlyar1998,
	title = {Addition spectrum, persistent current, and spin polarization in coupled quantum dot arrays: Coherence, correlation, and disorder},
	author = {Kotlyar, R. and Stafford, C. A. and Das Sarma, S.},
	journal = {Phys. Rev. B},
	volume = {58},
	issue = {7},
	pages = {3989--4013},
	numpages = {0},
	year = {1998},
	month = {Aug},
	publisher = {American Physical Society},
	doi = {10.1103/PhysRevB.58.3989},
	url = {https://link.aps.org/doi/10.1103/PhysRevB.58.3989}
}

@article{taylor2025neural,
  title={Neural network based deep learning analysis of semiconductor quantum dot qubits for automated control},
  author={Taylor, Jacob R and Das Sarma, Sankar},
  journal={Physical Review B},
  volume={111},
  number={3},
  pages={035301},
  year={2025},
  publisher={APS}
}

@article{white1992density,
  title={Density matrix formulation for quantum renormalization groups},
  author={White, Steven R},
  journal={Physical review letters},
  volume={69},
  number={19},
  pages={2863},
  year={1992},
  publisher={APS}
}

@article{schollwock2011density,
  title={The density-matrix renormalization group in the age of matrix product states},
  author={Schollw{\"o}ck, Ulrich},
  journal={Annals of physics},
  volume={326},
  number={1},
  pages={96--192},
  year={2011},
  publisher={Elsevier}
}

@inproceedings{hu2018squeeze,
  title={Squeeze-and-Excitation Networks},
  author={Hu, Jie and Shen, Li and Sun, Gang},
  booktitle={Proceedings of the IEEE/CVF Conference on Computer Vision and Pattern Recognition},
  pages={7132--7141},
  year={2018}
}

@article{moon2020machine,
  title = {Machine learning enables completely automatic tuning of a quantum device faster than human experts},
  author = {Moon, H. and Lennon, D. T. and Kirkpatrick, J. and van Esbroeck, N. M. and Camenzind, L. C. and Yu, L. and Vigneau, F. and Zumb\"uhl, D. M. and Briggs, G. A. D. and Osborne, M. A. and Sejdinovic, D. and Laird, E. A. and Ares, N.},
  journal = {Nature Communications},
  volume = {11},
  pages = {4161},
  year = {2020}
}

@article{taylor2025mitigating,
  title={Mitigating disorder and optimizing topological indicators with vision-transformer-based neural networks in Majorana nanowires},
  author={Taylor, Jacob R and Das Sarma, Sankar},
  journal={Physical Review B},
  volume={112},
  number={4},
  pages={L041110},
  year={2025},
  publisher={APS}
}

@article{itensor,
  title={{The ITensor Software Library for Tensor Network Calculations}},
  author={Fishman, Matthew and White, Steven R. and Stoudenmire, E. Miles},
  journal={SciPost Physics Codebases},
  pages={4},
  year={2022},
  doi={10.21468/SciPostPhysCodeb.4}
}

@article{dosovitskiy2020image,
  title={An image is worth 16x16 words: Transformers for image recognition at scale},
  author={Dosovitskiy, Alexey and Beyer, Lucas and Kolesnikov, Alexander and Weissenborn, Dirk and Zhai, Xiaohua and Unterthiner, Thomas and Dehghani, Mostafa and Minderer, Matthias and Heigold, Georg and Gelly, Sylvain and others},
  journal={arXiv preprint arXiv:2010.11929},
  url={https://arxiv.org/abs/2010.11929},
  year={2020}
}

@inproceedings{he2016deep,
  title={Deep residual learning for image recognition},
  author={He, Kaiming and Zhang, Xiangyu and Ren, Shaoqing and Sun, Jian},
  booktitle={Proceedings of the IEEE conference on computer vision and pattern recognition},
  pages={770--778},
  year={2016}
}

@article{taylor2026predicting,
  title={Predicting spin-orbit coupling in hole spin qubit arrays with vision-transformer-based neural networks on a generalized Hubbard model},
  author={Taylor, Jacob R and Laubscher, Katharina and Sarma, Sankar Das},
  journal={arXiv preprint arXiv:2604.05052},
  year={2026}
}

@article{stafford1994collective,
  title={Collective Coulomb blockade in an array of quantum dots: A Mott-Hubbard approach},
  author={Stafford, CA and Sarma, S Das},
  journal={Physical review letters},
  volume={72},
  number={22},
  pages={3590},
  year={1994},
  publisher={APS}
}

@article{sarma2001spin,
  title={Spin electronics and spin computation},
  author={Sarma, S Das and Fabian, Jaroslav and Hu, Xuedong and Zuti{\'c}, Igor},
  journal={Solid State Communications},
  volume={119},
  number={4-5},
  pages={207--215},
  year={2001},
  publisher={Elsevier}
}

@article{sarma2005spin,
  title={Spin quantum computation in silicon nanostructures},
  author={Sarma, S Das and de Sousa, Rogerio and Hu, Xuedong and Koiller, Belita},
  journal={Solid state communications},
  volume={133},
  number={11},
  pages={737--746},
  year={2005},
  publisher={Elsevier}
}

@article{durrer2020automated,
  title={Automated tuning of double quantum dots into specific charge states using neural networks},
  author={Durrer, Renato and Kratochwil, Benedikt and Koski, Jonne V and Landig, Andreas J and Reichl, Christian and Wegscheider, Werner and Ihn, Thomas and Greplova, Eliska},
  journal={Phys. Rev. Appl.},
  volume={13},
  number={5},
  pages={054019},
  year={2020},
  publisher={APS},
  url={https://journals.aps.org/prapplied/abstract/10.1103/PhysRevApplied.13.054019}
}

@article{wang2023automated,
  title={Automated Characterization of a Double Quantum Dot using the Hubbard Model},
  author={Wang, Will and Rooney, John Dean and Jiang, Hongwen},
  journal={arXiv preprint arXiv:2309.03400},
  year={2023}
}

@article{schuff2024fully,
   title={Fully autonomous tuning of a spin qubit},
   volume={9},
   ISSN={2520-1131},
   url={http://dx.doi.org/10.1038/s41928-025-01562-4},
   number={3},
   journal={Nature Electronics},
   publisher={Springer Science and Business Media LLC},
   author={Schuff, Jonas and Carballido, Miguel J. and Kotzagiannidis, Madeleine and Calvo, Juan Carlos and Caselli, Marco and Rawling, Jacob and Craig, David L. and van Straaten, Barnaby and Severin, Brandon and Fedele, Federico and Svab, Simon and Chevalier Kwon, Pierre and Eggli, Rafael S. and Patlatiuk, Taras and Korda, Nathan and Zumbühl, Dominik M. and Ares, Natalia},
   year={2026},
   month=feb, pages={304} }

@article{zwolak2024data,
  title={Data needs and challenges for quantum dot devices automation},
  author={Zwolak, Justyna P and Taylor, Jacob M and Andrews, Reed W and Benson, Jared and Bryant, Garnett W and Buterakos, Donovan and Chatterjee, Anasua and Das Sarma, Sankar and Eriksson, Mark A and Greplov{\'a}, Eli{\v{s}}ka and others},
  journal={npj Quantum Information},
  volume={10},
  number={1},
  pages={105},
  year={2024},
  publisher={Nature Publishing Group UK London}
}

@article{rao2025modular,
  title={Modular autonomous virtualization system for two-dimensional semiconductor quantum dot arrays},
  author={Rao, Anantha S and Buterakos, Donovan and van Straaten, Barnaby and John, Valentin and Yu, C{\'e}cile X and Oosterhout, Stefan D and Stehouwer, Lucas and Scappucci, Giordano and Veldhorst, Menno and Borsoi, Francesco and others},
  journal={Physical Review X},
  volume={15},
  number={2},
  pages={021034},
  year={2025},
  publisher={APS}
}

@article{buterakos2026qdflow,
  title={QDFlow: A Python package for physics simulations of quantum dot devices},
  author={Buterakos, Donovan L and Kalantre, Sandesh S and Ziegler, Joshua and Taylor, Jacob M and Zwolak, Justyna P},
  journal={SciPost Physics Codebases},
  pages={065},
  year={2026}
}
\clearpage
\renewcommand{\thesection}{\Roman{section}}
\vspace{3cm}
\onecolumngrid
\begin{center}
    {\bf \large Supplemental Materials for ``Large Scale Optimization of Quantum Dots through Tensor and Neural Networks''}
\vspace{1cm}
\end{center}
\twocolumngrid
\setcounter{page}{1}
\setcounter{secnumdepth}{3}
\setcounter{equation}{0}
\setcounter{figure}{0}
\renewcommand{\theequation}{S-\thesection.\arabic{equation}}

\renewcommand{\thefigure}{S\arabic{figure}}
\renewcommand\figurename{Supplementary Figure}
\renewcommand\tablename{Supplementary Table}

\section{Non-zero temperature process}
\label{sec:nonzerotemp}
To simulate the non-zero temperature effects one can perform the DMRG repeatedly but by adding a large energy cost to any state overlapping with the previously found ground state/excited states. The expectations in the non-zero temperature case are computed from the DMRG output $E_\lambda$ and states $\ket{\lambda}$ with the partial partition function $Z=\sum_\lambda e^{-\beta(E_\lambda-\sum_j\mu_j\bra{\lambda}n_j\ket{\lambda}}$ and inverse temperature $\beta$  as follows:

$$\vec{n}(\beta,\vec{\mu})_i=\sum_{\lambda} e^{-\beta(E_\lambda-\sum_j\mu_j\bra{\lambda} n_i\ket{\lambda})}\bra{\lambda} n_i\ket{\lambda}\frac{1}{Z}$$

Although we did attempt higher temperature effects, we found due to the need for a large dataset we were unable to afford to simulate the repeated DMRGs necessary at scale (though in principle this should scale approximately linearly with the number of states needed; our previous work found 12 more than sufficient but fewer may get approximate results too), even though our method did work for limited number of samples. Our current process only makes use of the ground state, where we thusly assume zero temperature, but this is not a limitation of our method. 

\begin{figure}[!t]
        \includegraphics[width=0.85\linewidth]{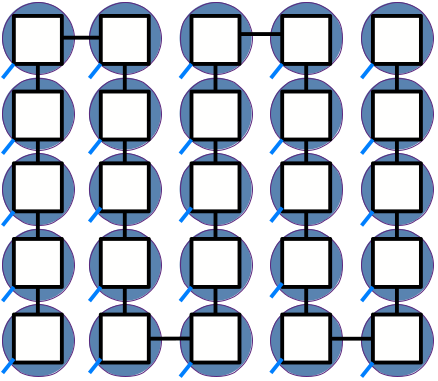}
    \caption{Tensor network diagram of the MPS representing the system's quantum state. The site connections snake back and forth in order to minimize long-range interactions and thereby reduce the required bond dimension. Tensors are represented as boxes, with their indices shown as lines. Free indices are unconnected lines (colored blue), which in this case correspond to the site indices, while inter-box lines (colored black) represent connected or shared indices.}\label{fig:TensorNetwork}
\end{figure}

\begin{figure}[!tbp]
    \centering
    \begin{subfigure}{0.49\linewidth}
    \centering\includegraphics[width=\textwidth]{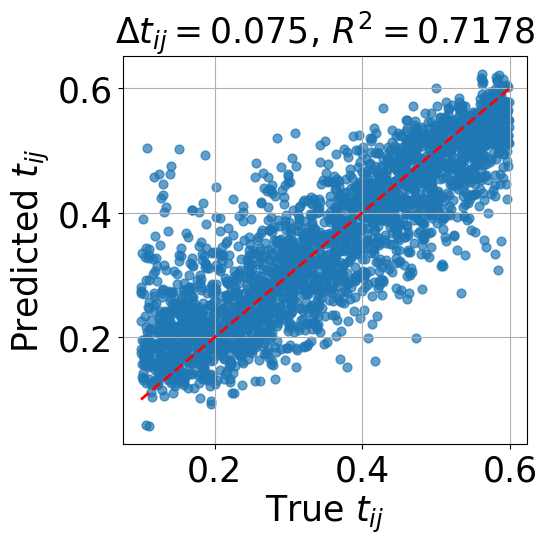}
        \caption{}
    \end{subfigure}
    \hfill
    \begin{subfigure}{0.49\linewidth}
        \centering
        \includegraphics[width=\textwidth]{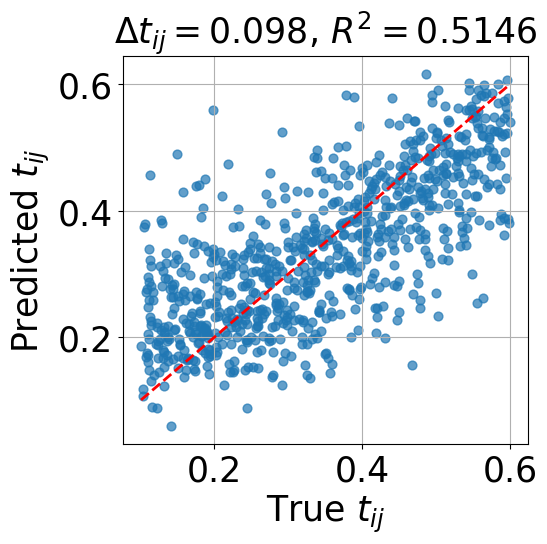}
        \caption{}
    \end{subfigure}
    \caption{Predicted vs. expected $t_{ij}$ for lattice grids of size (a) $3\times 3$ and (b) $5\times 5$. These results correspond to the case in which all parameters of the Hubbard model are unknown.}\label{fig:Allt}
\end{figure}

\begin{figure}[!tbp]
    \centering
    \begin{subfigure}{0.49\linewidth}
    \centering\includegraphics[width=\textwidth]{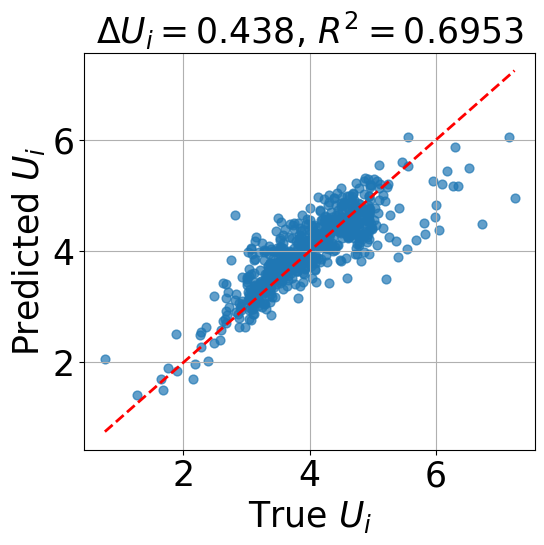}
        \caption{}
    \end{subfigure}
    \hfill
    \begin{subfigure}{0.49\linewidth}
        \centering
        \includegraphics[width=\textwidth]{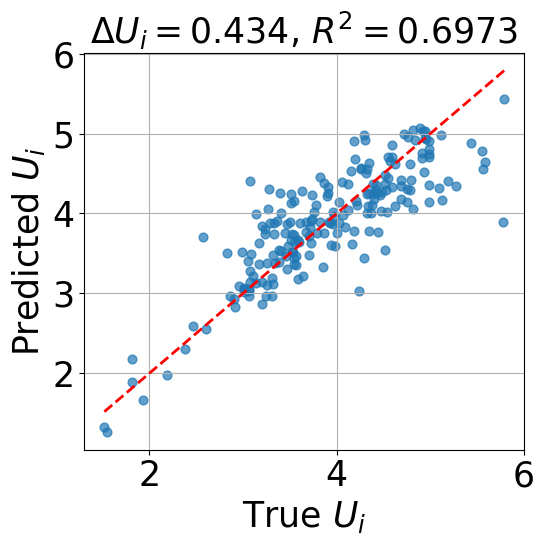}
        \caption{}
    \end{subfigure}
    \caption{Predicted vs. expected $U_i$ for lattice grids of size (a) $3\times 3$ and (b) $5\times 5$. These results correspond to the case in which all parameters of the Hubbard model are unknown.}\label{fig:AllU}
\end{figure} 
\begin{figure}[!t]
    \begin{subfigure}{0.49\linewidth}
    \centering\includegraphics[width=\textwidth]{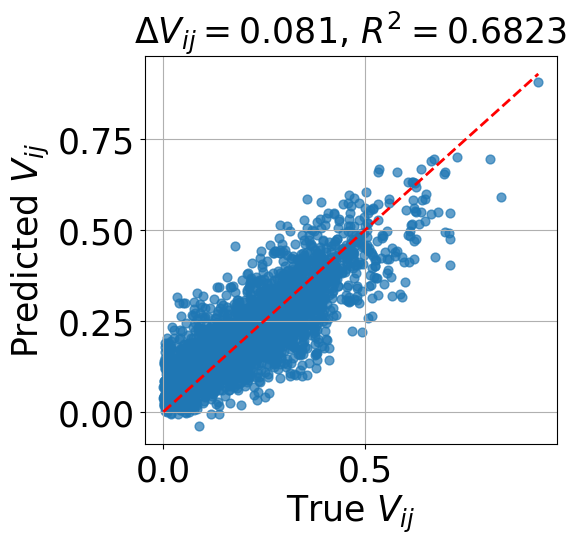}
        \caption{}
    \end{subfigure}
    \hfill
    \begin{subfigure}{0.48\linewidth}
        \includegraphics[width=\textwidth]{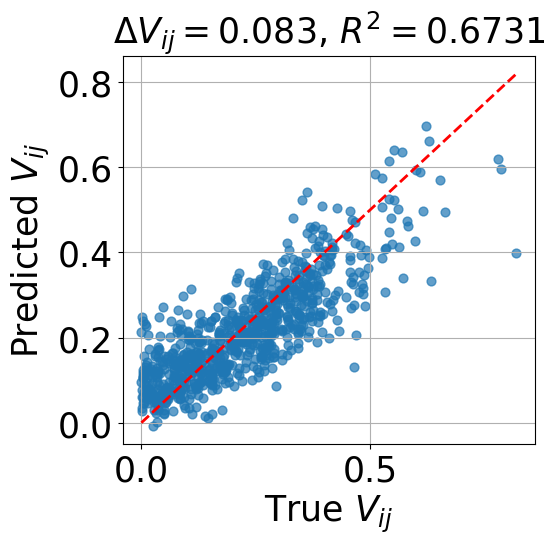}
        \caption{}
    \end{subfigure}
    \caption{Predicted vs. expected $V_{ij}$ for lattice grids of size (a) $3\times 3$ and (b) $5\times 5$. These results correspond to the case in which all parameters of the Hubbard model are unknown.}\label{fig:AllV}
\end{figure} 
\clearpage
\begin{figure*}[!tbp]
    \centering
    \includegraphics[width=0.99\textwidth]{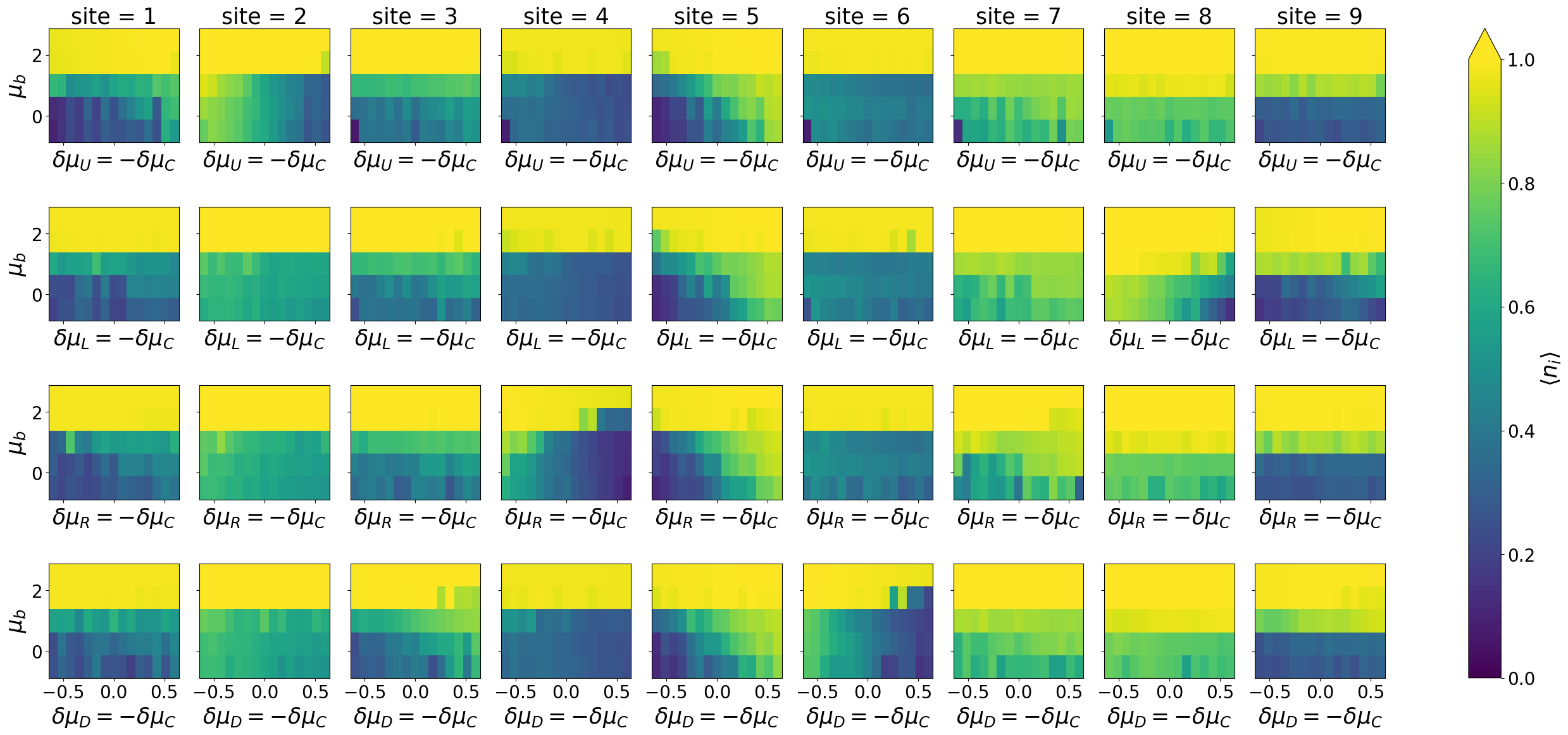}
    \caption{Example of Charge Stability Diagram used as input to the neural network. The occupation for $\langle n_i\rangle$ for $\mu_b$ vs. $\mu_c$ where the column is for site $i$ and row is for the center dots nearest neighbor link in the: Up (U), Left (L), Right (R) and Down (D) directions.}\label{fig:ChargeStability}
\end{figure*}

\begin{figure}[!t]
    \centering  
\includegraphics[width=0.6\linewidth]{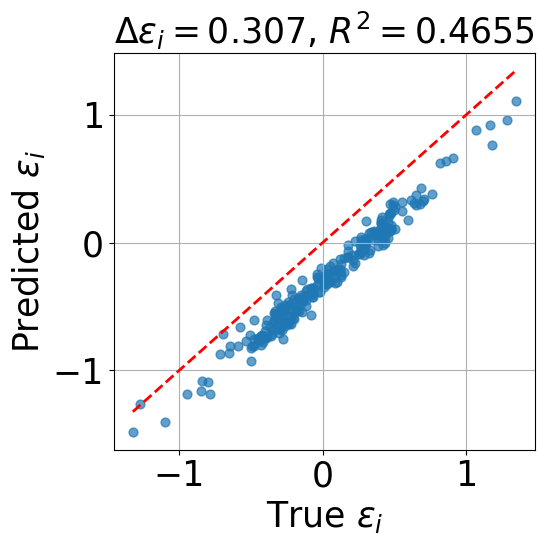}
    \caption{Prediction vs. expected $\epsilon_i$ for $5\times 5$ lattice grid when training with only $3\times3$ data without fine tuning. Here only on-site potential at all sites is treated as unknown. The neural network seems to be able to get relative magnitudes of on-site potential but with a linear displacement error.}\label{fig:NoFineTuning}
\end{figure}

\end{document}